\documentclass[useAMS,usenatbib]{mn2e}
\usepackage{amsmath}
\usepackage{txfonts}
\usepackage{graphics}
\usepackage{times}

\renewcommand{\vec}[1]{\mathbf{#1}}

\begin{document}


\title[Redshift space distortion models]{
Tests of redshift-space distortions models in configuration space for
the analysis of the BOSS final data release}
\author[White et al.]{
\parbox{\textwidth}{
    Martin White$^{1,2}$, Beth Reid$^{2}$,
    Chia-Hsun Chuang$^{3}$,
    Jeremy L. Tinker$^{4}$,
    Cameron K. McBride$^{5}$,
    Francisco Prada$^{3,6,7}$,
    Lado Samushia$^{8,9}$
}
\vspace*{4pt} \\
$^{1}$ Departments of Physics and Astronomy, University of California,
Berkeley, CA 94720, USA \\
$^{2}$ Lawrence Berkeley National Laboratory, 1 Cyclotron Road,
Berkeley, CA 94720, USA \\
$^{3}$ Instituto de F\'{\i}sica Te\'orica, (UAM/CSIC),
Universidad Aut\'onoma de Madrid,  Cantoblanco, E-28049 Madrid, Spain \\
$^{4}$ Center for Cosmology and Particle Physics, New York University,
New York, NY 10003, USA \\ 
$^{5}$ Harvard-Smithsonian Center for Astrophysics, 60 Garden St.,
Cambridge, MA 02138, USA \\
$^{6}$ Campus of International Excellence UAM+CSIC, Cantoblanco, E-28049
Madrid, Spain \\
$^{7}$ Instituto de Astrof\'{\i}sica de Andaluc\'{\i}a (CSIC),
Glorieta de la Astronom\'{\i}a, E-18080 Granada, Spain \\
$^{8}$ Department of Physics, Kansas State University, 116 Cardwell Hall,
Manhattan, KS 66506, USA \\
$^{9}$ National Abastumani Astrophysical Observatory, Ilia State University,
2A Kazbegi Ave., GE-1060 Tbilisi, Georgia
}
\date{\today}
\pagerange{\pageref{firstpage}--\pageref{lastpage}}

\maketitle

\label{firstpage}

\begin{abstract}
Observations of redshift-space distortions in spectroscopic galaxy surveys
offer an attractive method for observing the build-up of cosmological
structure, which depends both on the expansion rate of the Universe and our
theory of gravity.  In preparation for analysis of redshift-space distortions
from the Baryon Oscillation Spectroscopic Survey (BOSS) final data release
we compare a number of analytic and phenomenological models, specified in
configuration space, to mock catalogs derived in different ways from several
N-body simulations.
The galaxies in each mock catalog have properties similar to those of the
higher redshift galaxies measured by BOSS but differ in the details of how
small-scale velocities and halo occupancy are determined.
We find that all of the analytic models fit the simulations over a limited
range of scales while failing at small scales.  We discuss which models are
most robust and on which scales they return reliable estimates of the rate
of growth of structure: we find that models based on some form of resummation
can fit our N-body data for BOSS-like galaxies above $30\,h^{-1}$Mpc well
enough to return unbiased parameter estimates.
\end{abstract}


\section{Introduction}
\label{sec:intro}

The evolution of our Universe appears to be well described by Einstein's
theory of General Relativity.  The large-scale structure within it is
understood as the consequence of gravitational instability, which amplifies
primordial fluctuations laid down at early times.  Within this paradigm,
the growth of large-scale structure is driven by the motion of matter and
inhibited by the cosmological expansion.
Once the content, spatial geometry and initial perturbations in the Universe
are specified, Einstein's theory makes precise predictions for the expansion
rate and, simultaneously, the rate of growth of large-scale structure.
These predictions can be compared with observations made in galaxy redshift
surveys to test the theory and to provide constraints on the parameters of our
cosmological model.  We can measure the velocity field from maps of galaxies
in such surveys because the galaxy redshifts, from which distances are
inferred, include components from both the Hubble flow and peculiar
velocities from the comoving motions of galaxies
(see \citealp{HamiltonReview} for a review).
Thus even though the statistics of the galaxy distribution are isotropic,
redshift surveys exhibit an anisotropic distribution.  The anisotropy
encodes information about the build-up of structure and provides a sharp
test of the theory
\cite[see e.g.][for recent studies]{BerNarWei01,Zha07,Jain08,Guz08,NesPer08,
Song09a,Song09b,PerWhi08,McDSel09,WhiSonPer09,Song10,Zhao10,Song11}.

The measured anisotropy of the clustering of galaxies seen in redshift surveys
combines virial motions within halos on small-scales \citep{Jac72} and
supercluster-infall on large scales \citep{Kai87}.
Over the past two decades the measurement of these effects has become
increasingly precise
\citep{CFW95,Pea01,Per04,Ang08,Oku08,Guz08,WiggleZRSD,SamPerRac12,
Reid12,Samushia13,Bla13,Tor13,Samushia14,San14,Bel14,Toj14}.
Reliably extracting cosmological information from such precise measurements
requires models which are accurate, to the level of the data uncertainties,
on the scales over which they are fitted.
In this paper we compare how well a number of different models recover
the growth rate in mock catalogs made from N-body simulations, in order to
delineate the range of validity of the models in preparation for using them
to analyze data from the final release of data from the
Baryon Oscillation Spectroscopic Survey \citep[BOSS;][]{BOSS}, which
is part of Sloan Digital Sky Survey III \citep[SDSS-III;][]{Eis11}.
To this end we have developed a number of mock galaxy catalogs with
clustering properties similar to those of the higher redshift BOSS galaxies.
While having similar clustering and number density, the mock catalogs differ
in detail and have been generated to test various aspects of the
redshift-space distortion (RSD) models.

Our focus in this paper is on a set of analytic and phenomenological models
described in configuration space (and described further in
Section \ref{sec:models},
see also \citealt{OkuJin11,TorGuz12,Bia12,GM12} and references below).
Recent, complementary, tests of models in Fourier space have been presented in
\citet{JenBauPas11,WiggleZRSD,KwaLewLin12,OkuSelDes12,Zhe13,Oku14,Beutler14}
and N-body based models for fitting the data to smaller scales have been
presented in \citet{Reid14}.
The outline of the paper is as follows.  In Section \ref{sec:models} we
introduce the models we will be considering and the parameters upon which
they depend.  In Section \ref{sec:simulations} we describe the N-body
simulations and mock catalogs which we shall use as a reference for the
models to fit. Section \ref{sec:comparison} describes our methodology and
we discuss the implications in Section \ref{sec:discussion}.

\section{Models}
\label{sec:models}

In this section we describe the models which we shall fit to the
N-body data, with a focus on models which make predictions in
configuration space (recent, complementary tests of models in
Fourier space have been presented in the papers cited above).
The first two models are ``dispersion models'', which \citet{Scocc2004} showed 
inherently assume an unphysical pairwise velocity distribution.  The second two 
models express the redshift-space
correlation function as an integral of the real-space correlation function
times a pairwise velocity distribution function.

Throughout we adopt the standard ``plane-parallel'' or ``distant-observer''
approximation, in which the line-of-sight direction to each object is taken to
be the fixed direction $\hat{z}$. This has been shown to be a good
approximation at the level of current observational error bars
(e.g., Figure 10 of \citealt{SamPerRac12} or Figure 8 of \citealt{YooSel14}).

\subsection{The Eulerian dispersion model}

The simplest model for redshift space distortions combines the
supercluster infall enhancement of \citet{Kai87} with an independent
small-scale suppression to account for the virial motions of satellites in
halos (the finger-of-god effect).  In Fourier space the assumption
of an exponential pairwise velocity distribution with a scale-independent
width, along with the linear theory relation for supercluster infall, leads
to \citep{Pea92,Park94,PD94,BalPeaHea96,HamiltonReview,HatCol99}:
\begin{equation}
  P(k,\mu) = P_r(k)\frac{(b+f\mu^2)^2}{1+k^2\mu^2\sigma_E^2}
\label{eqn:esm}
\end{equation}
where $\sigma_E$ is a free parameter to be fit to the data, $b$ is the
large-scale (assumed scale-independent) bias and
$f=d\ln\delta/d\ln a\approx \Omega_m^{0.55}(z)$
is the growth rate of perturbations.
A motivation for the exponential form of the pairwise velocity distribution
can be found in \citet{Whi01,Sel01}, expressions for the Legendre moments
of Eq.~(\ref{eqn:esm}) can be found in \citet{CFW95}.
The exact form of the small-scale suppression, and the value of $\sigma_E$,
are strongly dependent on the galaxy population being modeled
\citep{JinBor04,Li07}.
Sometimes a model with $(1+k^2\mu^2\sigma_E^2/2)^2$ in the denominator is used;
this agrees with Eq.~(\ref{eqn:esm}) for low $k$.
An alternative formulation is to take the high-$k$ damping term as
a Gaussian, $\exp[-k^2\mu^2\sigma_E^2]$.
Again the expressions match for low $k$.

In Eq.~(\ref{eqn:esm}) we can use the linear theory power spectrum, or we
can use a model for the non-linear power spectrum (but ignoring non-linear
bias and the generic perturbation theory differences in the non-linear
evolution of velocity and density).
In the study of \citet{WiggleZRSD}, Eq.~(\ref{eqn:esm}) with a non-linear
power spectrum calculated as in \citet{HaloFit} performed very well against
data when fitted in Fourier space.
We shall consider both the linear and non-linear forms, referring to the
former as the Eulerian Dispersion Model (EDM) and the latter as the
Non-linear Dispersion Model (NDM) \citep[following][]{Pee80,Fis95}.
\citet{TinWeiZhe06} and \citet{Tin07}, which are developments of the work
in \citet{HatCol99}, discuss a number of improvements to models like this
based on fits to numerical simulation results.  We shall not consider these
improvements, since the original form is the most widely used.

The correlation function, which shall be the primary focus of this paper,
is the Fourier transform of this power spectrum.  If we expand the
correlation function in Legendre polynomials, $L_\ell$,
\begin{equation}
  \xi(s,\widehat{s}\cdot\widehat{z}) =
  \sum_\ell \xi_\ell(s)L_\ell(\widehat{s}\cdot\widehat{z})
\end{equation}
where $\vec{s}$ is the redshift-space coordinate, then
\begin{equation}
  \xi_\ell(s) = (2\ell+1) i^\ell \int \frac{d^3k}{(2\pi)^3}
  P(k,\mu)L_\ell(\mu)j_\ell(ks)
\end{equation}
where $j_\ell$ is a spherical Bessel function of order $\ell$.
We shall restrict our attention to the lowest non-vanishing moments,
$\ell=0$ and 2.

This model has three free parameters, which we can take to be either
$b\sigma_8$, $f\sigma_8$ and $\sigma_E$ or
$b$, $f$ and $\sigma_E$ depending on whether we wish to treat the linear
theory power spectrum amplitude as known.

We note that the model of \citet{Chuang13} is almost the same as the EDM,
differing only in the modeling of the acoustic peak region.  Since the
acoustic feature has very little weight in our fits we expect our
discussion of the EDM will apply approximately to this model as well.

\subsection{Perturbation theory inspired model}

We also consider a model inspired by, but not directly derived from,
perturbation theory in combination with the reasoning described above.
We call this model the perturbation-inspired model (PIM).
It was used to interpret the BOSS data in \citet{San13,San14}.
This model assumes
\begin{equation}
  P(k,\mu)=\left(\frac{b+f\mu^2}{1+k^2f^2\sigma_P^2\mu^2}\right)^2
  \left[ e^{-k^2\sigma_P^2}P_{\rm lin}(k)+A\,P_{22}(k) \right]
\end{equation}
where $b$, $f$, $\sigma_P$ and $A$ are free parameters and $P_{22}$ is the
standard second-order, mode-coupling term in Eulerian perturbation theory
\citep{Pee80,Jus81,Vis83,GGRW86,MSS92,JaiBer94}.
For further details on, and tests of, the model see \citet{San13,San14}.
In comparison with the other models note the factors of $f$ in the prefactor,
the squaring of the finger-of-god term, the double duty played by the
parameter $\sigma_P$ and the isotropic nature of the broadening of the
acoustic peak (i.e.~the exponential multiplying $P_{\rm lin}$).
As it stands this model has one additional parameter ($A$) that can be varied
in the fit beyond the large-scale bias, $b$ and $\sigma_P$.

\subsection{The Gaussian streaming model}

The ``Gaussian streaming model'' (GSM) was developed in \citet{ReiWhi11,Reid12},
where it was shown to fit the monopole and quadrupole of the correlation
functions of mock galaxies with a large-scale bias $b\simeq 2$ to the per
cent level on scales above $25\,h^{-1}$Mpc.  This model has been used to
interpret the clustering of galaxies measured in BOSS by
\citet{Reid12,Samushia13,Samushia14}.
The GSM is inspired by the work of \citet{Pee80,Fis95}.
The pairwise velocity distribution is different for components along and
perpendicular to the (real space) pair separation vector $r\widehat{r}$,
with the mean pairwise velocity $v_{12}(r) \widehat{r}$ oriented along
$\widehat{r}$.  The mean and variance of the pairwise velocity
distribution for the line-of-sight (LOS) velocity component therefore depends on
$\mu_r = \widehat{r} \cdot \widehat{z}$.  The GSM approximates the pairwise
velocity distribution as Gaussian and enforces pair conservation by integrating
over all possible real space LOS pair separations $y$ that appear at redshift
space separation $s_{\parallel}$.  Specifically we assume that the
redshift-space halo correlation function is
\begin{equation}
  1 + \xi^s(s_\perp, s_\parallel) =
  \int \dfrac{dy}{\sqrt{2\pi}\sigma_{12}}
  [1 + \xi(r)]
  \exp\left\{-\frac{[s_\parallel-y-\mu_r v_{12}]^2}{2\sigma_{12}^2}\right\}\ ,
\label{eqn:gsm}
\end{equation}
where $\xi(r)$, $v_{12}$ and $\sigma_{12}$ are to be provided from an
analytic theory.  In its basic form \citep{ReiWhi11,Reid12} integrated
Lagrangian perturbation theory with scale-dependent but local Lagrangian bias
\citep{Mat08b} is used for the real-space correlation function of halos,
while the halo infall velocity and dispersion computed in standard
perturbation theory with scale-independent bias.
In order to go from halos to galaxies, \citet{Reid12} showed that it suffices
to introduce a single additional parameter, $\sigma_{FOG}$, akin to the
$\sigma_E$ in Eq.~(\ref{eqn:esm}).
This is taken to be an isotropic, scale-independent dispersion which is
added in quadrature to $\sigma_{12}$ and modifies the scale-dependence of the
quadrupole moment on small scales.

\subsection{The Lagrangian streaming model}

The Lagrangian streaming model (LSM) is a hybrid model based upon the work
in \citet{ReiWhi11,Reid12,CLPT,WanReiWhi13,ZSM}.
It uses  the Zel'dovich approximation \citep{Zel70} with a model for local
Lagrangian bias introduced by \citet{Mat08a,Mat08b,Mat11} to predict the
real-space correlation function of biased tracers.
This model predicts the real-space statistics very well, compared to
N-body simulations, but fares less well for the redshift-space statistics of
biased tracers, particularly the quadrupoles.  For this reason the real-space
correlation function is ``convolved'' with a Gaussian as in
Eq.~(\ref{eqn:gsm}), except that the dispersion is computed from Lagrangian
rather than Eulerian perturbation theory and we shall denote it as $\sigma_L$.
As discussed extensively in \citet{ReiWhi11}, the measured quadrupole is
very sensitive to the pairwise velocity, $v_{12}$.  For this reason we use
the 1-loop expansion for $v_{12}$ rather than just the lowest order
(i.e.~Zeldovich) term.
[Comparison of this model with the full 1-loop calculation in
\citet{WanReiWhi13} indicates that the differences are at the percent level
on scales larger than $20\,h^{-1}$Mpc, except around the acoustic peak.
Since the acoustic peak carries little weight in our fits we can treat the
LSM and the model described in \citet{WanReiWhi13} as essentially identical,
though the LSM is computationally easier.]
As in the GSM, an additional parameter needs to be included to model
fingers-of-god.  We considered two approaches: an additional, isotropic,
component to the Gaussian as in the GSM and an exponential form for the
extra dispersion.  In general we find that the exponential form produces
almost identical results to the Gaussian form.
Since the latter is easier to implement we shall use it throughout.

In principle the LSM depends on several parameters, but we choose to fix the
bias parameters using the peak-background split expressions as described in
\citet{ZSM}.  We have found that allowing the second bias parameter to float
in the fits produces almost the same correlation functions as fixing it to the
peak-background expression, so we work with the reduced parameter space.
For a fixed linear power spectrum then, the final model depends on a single
bias parameter, $f$ and $\sigma_L$; i.e.~the same number of parameters as the
Eulerian model above.

\subsection{Comparison}

\begin{figure}
\begin{center}
\resizebox{3.3in}{!}{\includegraphics{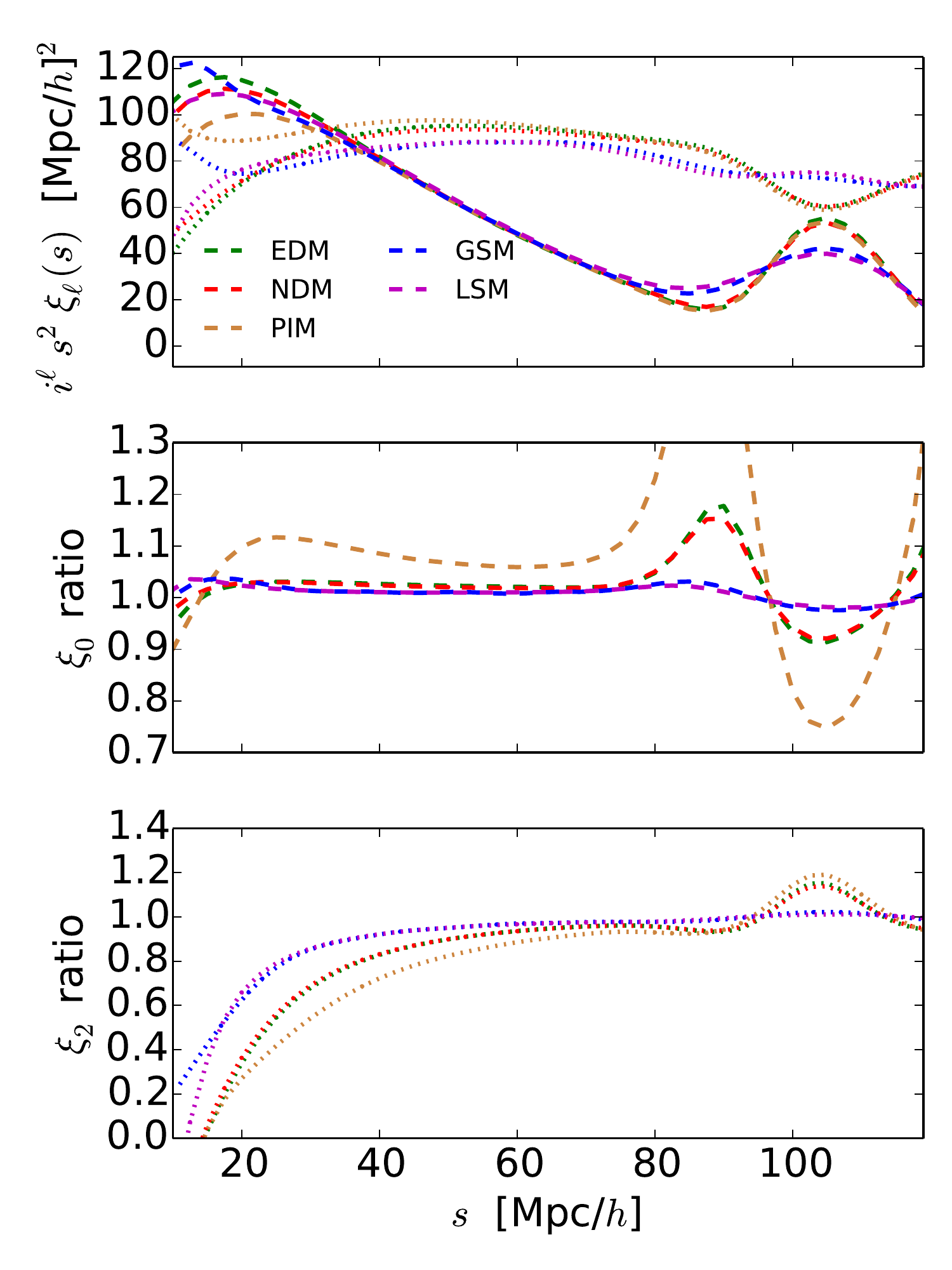}}
\caption{A comparison of the theoretical models described in Section
\ref{sec:models}.  In the top panel the predictions for the monopole moment
of the correlation function (multiplied by $s^2$ to reduce the dynamic range)
are shown as dashed lines while those for the quadrupole moment are shown as
dotted lines.  We have taken the finger-of-god parameter, $\sigma$, to be
small ($0.5\,h^{-1}$Mpc) to highlight the differences in the perturbative part
of the models.  In the lower panels we plot the ratio of the monopole and
quadrupole moments with $\sigma=5\,h^{-1}$Mpc to those with
$\sigma=0.5\,h^{-1}$Mpc to show the effects of varying $\sigma$.
Generally, higher $\sigma$ goes with a sharper ``break'' at small $s$ in the
quadrupole and lower clustering at small $s$ in the monopole but $\sigma$
has a much larger effect on the quadrupole than the monopole, and alters small
scales more than large.
For such highly biased galaxies the effect of the finger-of-god modeling
becomes comparable to the observational uncertainty for BOSS-like surveys
below about $30\,h^{-1}$Mpc.
The broadening of the acoustic peak in the Lagrangian models compared to the
Eulerian models is clearly evident.
For the perturbation theory inspired model (PIM) we have set $A=1$.}
\label{fig:theory}
\end{center}
\end{figure}

Figure \ref{fig:theory} compares the monopole and quadrupole moments of the
redshift-space correlation function of halos for the different theoretical
models described above.  We used the linear theory power spectrum appropriate
to the T1 simulations (see Section 3) and a large-scale bias of $b=2$ for all
models, and have varied the finger-of-god parameters, $\sigma_i$, from
$0.5\,h^{-1}$Mpc to $5\,h^{-1}$Mpc to illustrate their effects.
Note that the $\sigma_i$ have a much larger effect on the quadrupole than the
monopole, altering small scales more than large.
For such highly biased galaxies the effect of the finger-of-god modeling
becomes comparable to the observational uncertainty for BOSS-like surveys
below about $30\,h^{-1}$Mpc.
The broadening of the acoustic peak in the Lagrangian models compared to the
Eulerian models is clearly evident.

\section{Simulations}
\label{sec:simulations}

\begin{table}
\begin{center}
\begin{tabular}{ccccccc}
Name    & $\Omega_m$ &  $h$  & $f^{(0.55)}$  & $\sigma_8^{(0.55)}$ &
  $\bar{n}$ & Vol  \\
\hline
T0      & 0.274      & 0.70  & 0.74 &  0.61      & 3.6 &  68   \\
T1      & 0.292      & 0.69  & 0.76 &  0.62      &  4  &  26   \\
MD      & 0.307      & 0.68  & 0.77 &  0.61      &  4  &  16   \\
QPM     & 0.290      & 0.70  & 0.76 &  0.61      & 4.1 & 268
\end{tabular}
\end{center}
\caption{A summary of information about the simulations used in this paper
(see text for further discussion).
Each is of the $\Lambda$CDM family with the indicated values of the matter
density ($\Omega_m$) and Hubble parameter ($h$).  We use a single redshift
output to build the mock catalogs, which is near $z=0.55$.  The growth rate
at this redshift, $f=d\ln D/d\ln a\approx \Omega_m^{0.55}$ where $D(a)$ is
the linear theory growth factor.  The normalization of the linear theory
power spectrum at $z=0.55$ is given in terms of $\sigma_8(z=0.55)$, denoted
$\sigma_8^{(0.55)}$.
Finally we give the average number density, in $10^{-4}\,h^{-3}{\rm Mpc}^3$,
and the total volume, in $h^{-3}{\rm Gpc}^3$, used in the computation of
$\xi_\ell$.  For model T1 we have three variants (00, 01 and 10) which are
described further in the text.}
\label{tab:sims}
\end{table}

In order to validate these different models we use mock catalogs derived from
N-body simulations.  Such validations have been performed for many of the
models, often in the paper in which they were introduced, however our goal is
to enable a side-by-side comparison using a common set of simulations, and to
test the models for mock galaxies with properties similar to those of BOSS
galaxies at $z\simeq 0.5$.  We have created these mock catalogs from a number
of different simulations, with differing cosmologies, codes and mock catalog
generation methods (see Table \ref{tab:sims} for a summary).  While each is
comparable to the BOSS data, the differing assumptions provide a good check
of the models' ability to handle the complexities of how real galaxies trace
the cosmic web.

\subsection{TreePM}
\label{sec:treepm}

The first set of simulations which we use (labeled T0 below) are those
used in \citet{Whi11,ReiWhi11,WanReiWhi13,ZSM}.
The catalogs are based on twenty realizations of the $\Lambda$CDM model
with $\Omega_m=0.274$ and $h=0.7$, each employing $1500^3$ particles in a
periodic box of side length $1500\,h^{-1}$Mpc for a total volume of
$68\,h^{-1}{\rm Gpc}^3$.
The simulations were run with the TreePM code of \citet{TreePM} and the
mock catalogs are described further in \citet{Whi11}.
Briefly, halos were found using the friends-of-friends algorithm and
populated with galaxies resembling those of BOSS using a halo occupation
distribution (HOD).
Each central galaxy was placed at the minimum of the halo potential and
given a velocity equal to that of the halo center-of-mass, while halo
particles were picked at random to model satellite galaxies.

A second set of simulations (labeled T1), run with the same code, is
also used (these simulations have also been used in \citealt{Reid14}).
This set has less volume but higher mass and force resolution and a different
cosmology.  This set consists of ten realizations of a $\Lambda$CDM model
with $\Omega_m=0.292$ and $h=0.69$, each employing $2048^3$ particles in a
periodic box of side length $1380\,h^{-1}$Mpc.  At each output of the
simulation, two sets of halo catalogs were constructed.  The first is based
on the friends-of-friends algorithm as above.  The second uses a spherical
overdensity criterion.  As before a HOD was used to generate mock galaxies.
For the friends-of-friends halos, central and satellite galaxies were placed
in the halo as above, except for one of the simulations (labeled 01) the
satellite velocities were boosted by 25 per cent while in the base simulation
(labeled 00) they were not.
For the spherical overdensity catalog (labeled 10) the central galaxies were
placed at rest compared to the inner region of the halo, rather than the halo
center-of-mass.  Since massive halos contain significant kinematic
substructure, these prescriptions can differ by a fair fraction of the halo
velocity dispersion, and generate differing amount of small-scale
redshift-space distortions.  Taken together these simulations explore a
range of cosmologies and galaxy velocity prescriptions, as will be seen in
the clustering statistics.  We generated some other variants of these models,
but in all cases the results were consistent with those we explore in more
detail below so we focus on the models listed from now on.

Especially for the higher resolution simulations, the box size is not optimal
for resolving features (such as the baryon acoustic peak) at $>100\,h^{-1}$Mpc.
Since $100\,h^{-1}$Mpc is about 10 per cent of the side length of the periodic
box the missing long-wavelength modes have an impact on the correlation
function and the large-scale velocities which act to broaden the acoustic peak.
Fortunately, most of the RSD signal comes from smaller scales so this is not
a major concern for the purposes of this work.

Due to the combination of large volume and high force and mass resolution
we shall take the T1  simulations as our fiducial choice in the figures below,
and we will comment specifically when the results from the other simulations
show a qualitatively different behavior.
We shall also use the SO catalog (denoted 10 above) as our fiducial model, as
this catalog provides a very good match to the projected correlation function
and monopole and quadrupole moments of the redshift-space correlation function
of CMASS galaxies in BOSS \citep{Reid14}.

\subsection{Big MultiDark}
\label{sec:bigmd}

A further set of mock catalogs were generated from the ``Big MultiDark''
(hereafter BigMD or MD) simulation.
This simulation employed $3840^3$ particles in a periodic box of side
length $2500\,h^{-1}$Mpc.  The cosmology was close to the best-fitting
Planck cosmology \citep{Planck1}, with $\Omega_m=0.307$.
Unlike the halo occupation based methods described above, these catalogs
were produced by halo abundance matching as described in \citet{Nuza13},
with galaxies associated to substructures.
This has a more complex, and possibly realistic, velocity distribution than
the HOD models above.
Unfortunately the total volume of this simulation is only slightly larger
than our fiducial mock survey (see below) so significant finite-volume `noise'
remains.

\subsection{Quick particle mesh}

Later we will use a set of {\it approximate\/} mock catalogs, which were run
in order to generate covariance matrices for the BOSS measurements via
Monte-Carlo techniques \citep[see][for further discussion]{QPM}.
These mock catalogs use low mass and force resolution particle-mesh
simulations employing $1280^3$ particles in a $2560\,h^{-1}$Mpc box run
with large time steps.  At select times the particles, and their local
density smoothed on $2\,h^{-1}$Mpc scales, were dumped and these particles
were then sampled (with a density-dependent probability) to form a set of
mock halos which are then populated using a halo occupation distribution
\citep{QPM}.
These approximate mock catalogs contain roughly the right amount of monopole
and quadrupole power, but deviate from the high resolution simulations on
scales smaller than $30-50\,h^{-1}$Mpc.
We shall use these very large volume simulations, in a differential manner,
to test the impact of observing strategy on the recovered cosmological
parameters.

\begin{figure}
\begin{center}
\resizebox{3.3in}{!}{\includegraphics{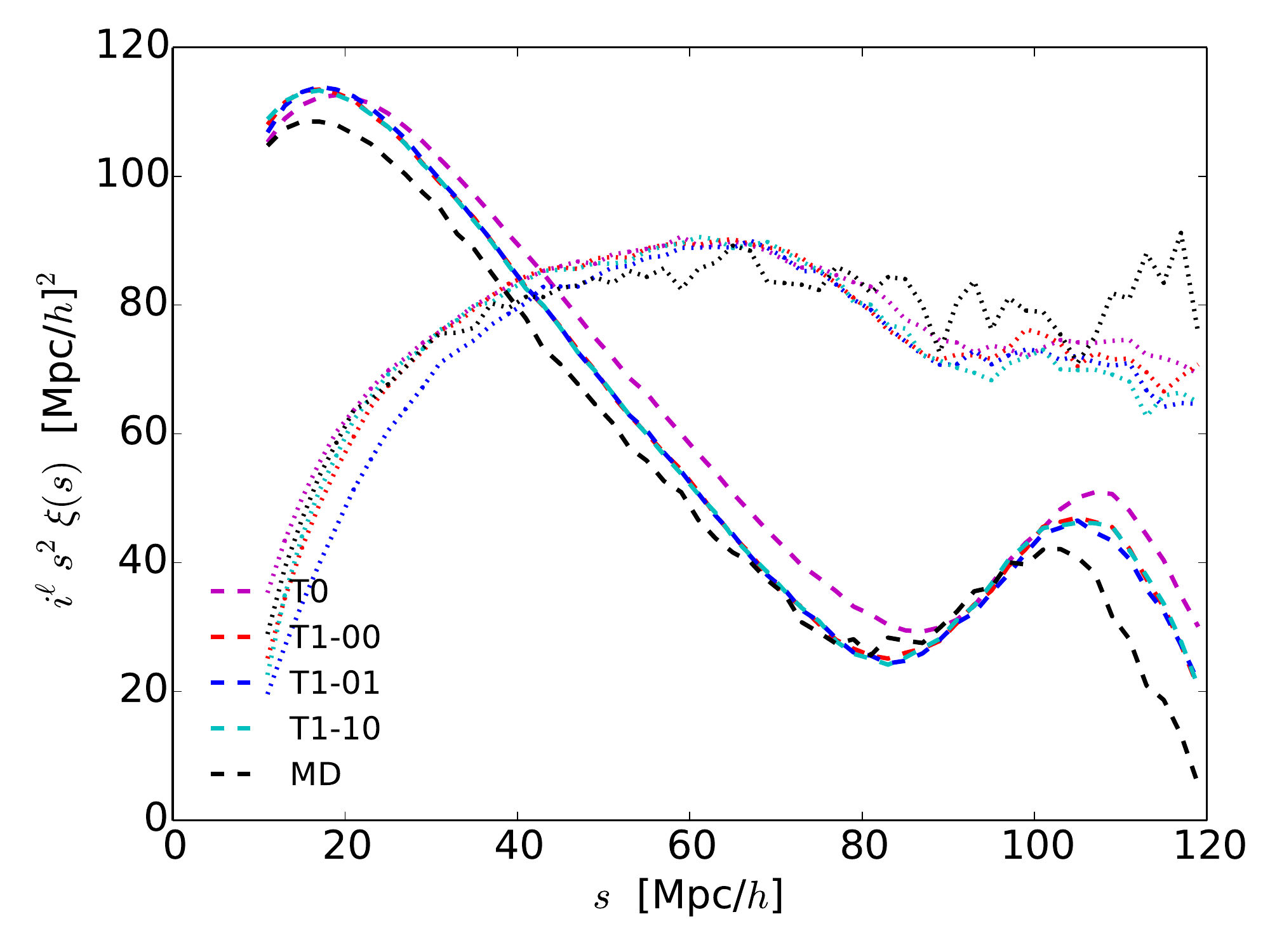}}
\caption{The monopole and quadrupole moments of the correlation function,
times $s^2$ to reduce the dynamic range in the plot, for the simulations
described in the text (see also Table \ref{tab:sims}).
In each case the mean value of the statistic is plotted, in bins of width
$2\,h^{-1}$Mpc, and error bars have been suppressed for clarity.
Models T0 and T1 are described in section \ref{sec:treepm} while MD is
described in section \ref{sec:bigmd}.  Model T1 has three variants, which
differ in terms of the halo finder and whether a velocity bias is assumed
for the satellites.}
\label{fig:moments}
\end{center}
\end{figure}

\section{Comparison}
\label{sec:comparison}

\subsection{Ideal data}

We begin by asking to what extent each of our models can recover the
simulation parameters when applied to ``ideal'' data.  To this end
we generate correlation functions from fixed-time outputs of the simulations
assuming no observational non-idealities.  For the TreePM and QPM runs we
use the periodic outputs and generate a mean and variance for each point from
the independent runs.  For the BigMD run, we divide the $2500\,h^{-1}$Mpc box
into its 8 octants and compute the correlation function in each octant, using
the octants to determine the mean and variation on the parameters.

We show the monopole and quadrupole moments of the mock galaxies for a
selection of the catalogs in Fig.~\ref{fig:moments}.  Note that the small
scale behavior of the quadrupole moment depends upon the model chosen for
the satellite and central velocities, while the amplitude of the monopole
moment is largely set by mimicking the BOSS data.
The range of slopes and behaviors reflects the differences in the underlying
cosmology, linear power spectrum and bias prescription and makes for a good
test of the theoretical models.

\begin{figure}
\begin{center}
\resizebox{3.3in}{!}{\includegraphics{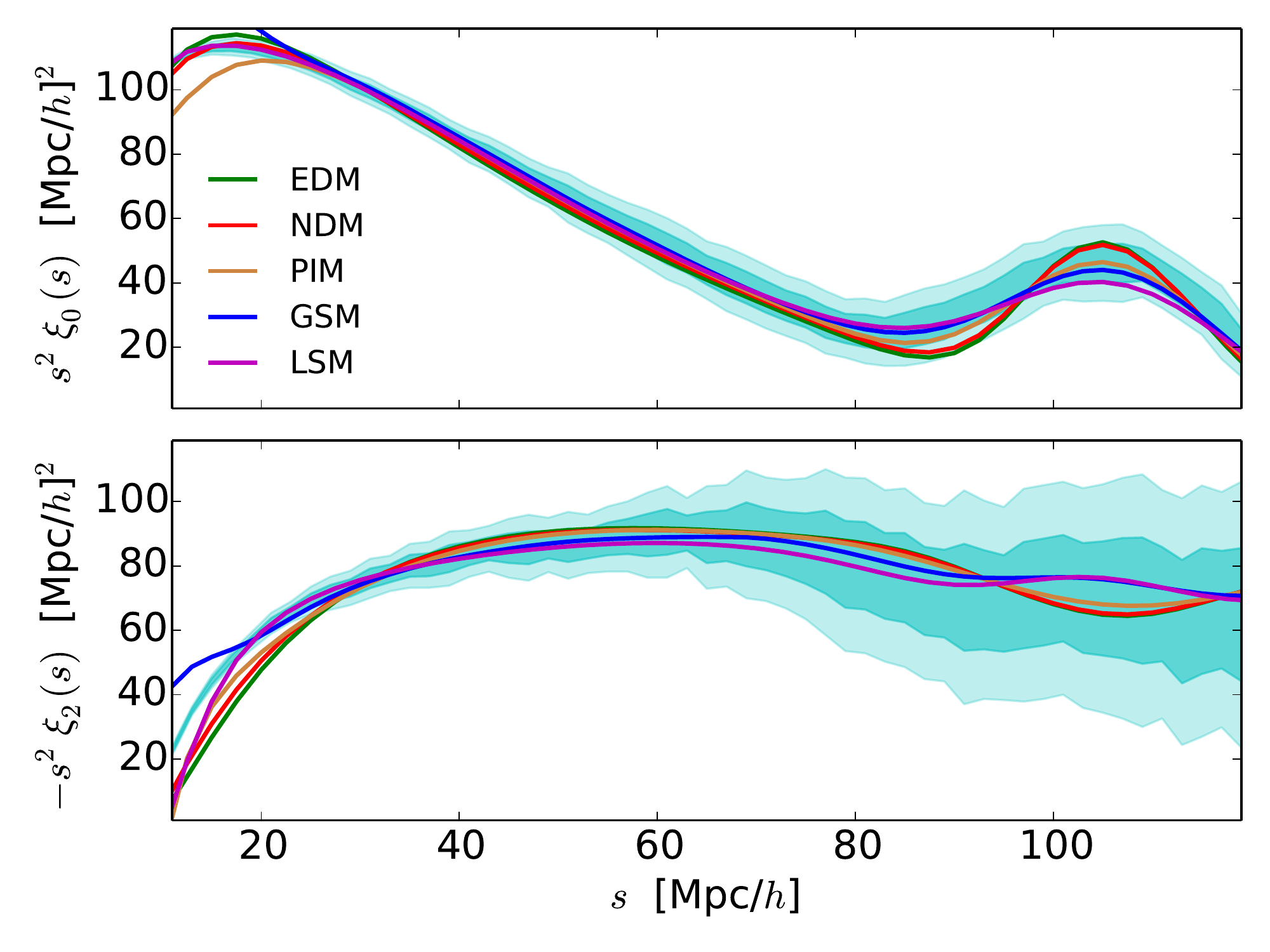}}
\caption{The best-fit models to the T1 simulation at $z\simeq 0.55$.
Each model was fit to the data in the range $30<s<120\,h^{-1}$Mpc, holding
the linear theory power spectrum and $f$ fixed at the appropriate values for
this simulation.  The shaded bands around the simulation results indicate the
adopted $1$ and $2\,\sigma$ errors, derived from the diagonal elements of the
covariance matrix.
The error bars are highly correlated.  The sharper acoustic peak in the
monopole of the Eulerian models than the Lagrangian models is expected,
as the Eulerian models do not account for non-linear broadening of the peak.
Some of the modes which are responsible for this broadening are not captured
by the T1 simulations, so the N-body peak is not as broad as it should be.
The models are quite similar in their monopole predictions at intermediate
scales however and the large scales carry little weight in the fit.}
\label{fig:bestfitPB}
\end{center}
\end{figure}

\begin{figure}
\begin{center}
\resizebox{3.3in}{!}{\includegraphics{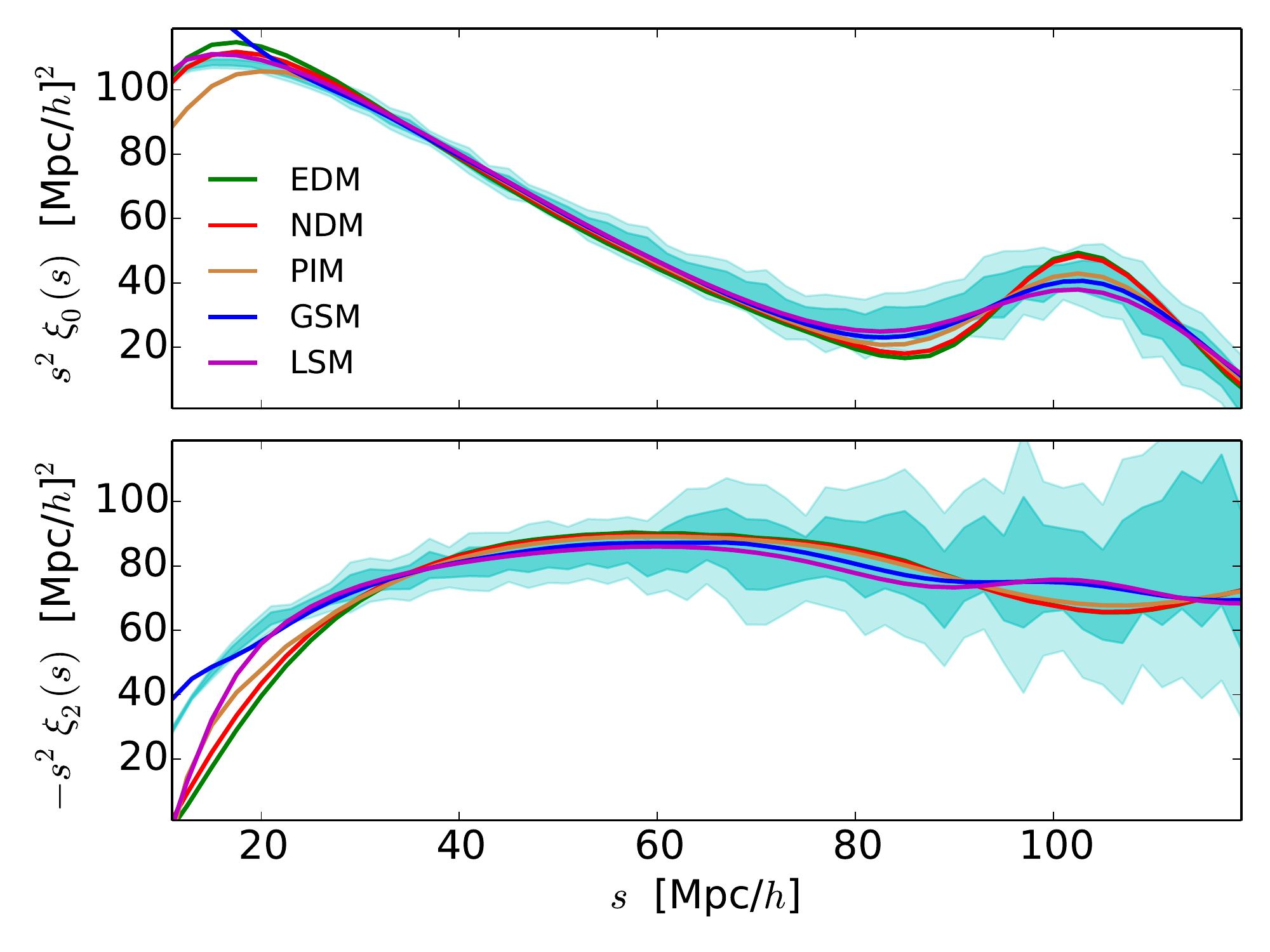}}
\caption{As in Fig.~\ref{fig:bestfitPB} but for the BigMD simulation at
$z\simeq 0.56$.  The error bars are highly correlated and the noise from
the smaller total volume is apparent.}
\label{fig:bestfitMD}
\end{center}
\end{figure}

\begin{figure*}
\begin{center}
\resizebox{6.5in}{!}{\includegraphics{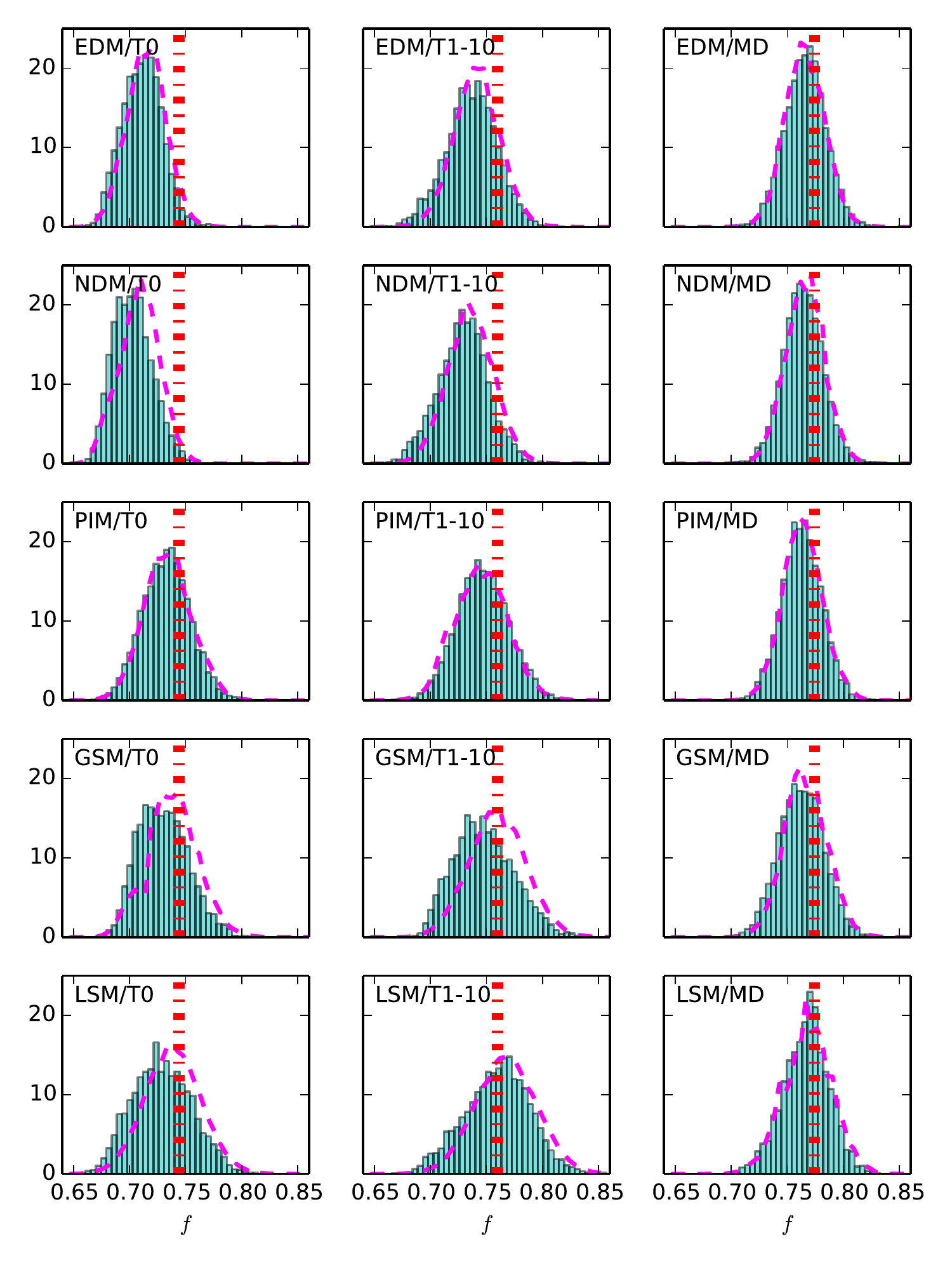}}
\caption{The profile likelihood (dashed lines) and the marginalized posterior
distribution of $f$ (filled histogram; from a Monte-Carlo Markov Chain), for
the models described in Section \ref{sec:models} and the simulations described
in Section \ref{sec:simulations}.  Scales larger than $30\,h^{-1}$Mpc have
been fit, as described in the text.
The vertical line marks the value of $f$ appropriate to each cosmology.}
\label{fig:fdist}
\end{center}
\end{figure*}

In order to provide fits to the data we need a covariance matrix.  We
use the linear theory expressions\footnote{The formula given assumes
infinitesimal bins.  We additionaly account for the finite bin width
by integrating $s^2\,j_\ell$ over the bin.}
\citep{Ber94,EisZal01,Coh06,Huf07}
\begin{eqnarray}
  {\rm Cov}\left[\xi_{\ell_1}(s_1),\xi_{\ell_2}(s_2)\right] &=&
  \frac{2}{V}i^{\ell_1+\ell_2}(2\ell_1+1)(2\ell_2+1)
  \int \frac{d^3k}{(2\pi)^3} \nonumber \\
  &\times& P^2(k,\mu) L_{\ell_1}(\mu) L_{\ell_2}(\mu)
  j_{\ell_1}(ks_1)j_{\ell_2}(ks_2)
\end{eqnarray}
in order to generate a correlation matrix (sometimes referred to as a
reduced covariance matrix).  Comparison of the linear theory correlation
matrix with that produced from the QPM mock catalogs shows they are quite
similar in structure \citep[see also][]{Reid12}.
The correlation matrix is independent of the volume of the survey or any
scaling of the bins (e.g.~whether it is $s^2\xi_0$ or $\xi_0$ that is being
constrained).
The covariance matrix we use is then obtained from the linear theory
correlation matrix by multiplying each row and column of the correlation
matrix by the standard deviation per bin, measured from the independent
simulations or octants, scaled (with error $\propto V^{-1/2}$) as if for a
$5\,(h^{-1}{\rm Gpc})^3$ volume, roughly the volume expected for high-$z$
galaxies in the final BOSS dataset.
This ensures that we can assess sub-$\sigma$ systematics with respect to the
BOSS data but we are not simply fitting noise from the finite number of mock
catalogs (since the volume of the mocks exceeds $5\,(h^{-1}{\rm Gpc})^3$ in
all cases).  By matching the error bars to those expected from BOSS we can
also assess how degeneracies and nuisance parameters affect the fits in a way
relevant to BOSS.
In preparing these matrices and $\xi_\ell$ measurements we cover
$10<s<120\,h^{-1}$Mpc using $2\,h^{-1}$Mpc bins\footnote{We have tested that
increasing the bin width to $4\,h^{-1}$Mpc does not alter the results.}.
We work with the first two even  multipole moments, $\xi_0$ and $\xi_2$.
In order to drop points at low or high $s$ we simply add a large number to
the nominal error bar before computing the covariance matrix and its inverse.
This provides a simple way of restricting the fitting range which requires
minimal changes to the code (though it does not fully remove the influence
of the small scale points, which requires zeroing the correlations, but when
the model is close to the data the difference is small and the approximation
does not qualitatively affect our conclusions).

We begin by fitting the models with the ``correct'' linear theory
power spectrum and cosmology (although we allow $f$ to float in order to see
if we recover the correct value).  Since the distance scale and the shape of
the power spectrum are not allowed to vary this test is quite constraining.
There is no ``slop'' from other parts of the theory to hide a bad fit.
Our nominal fitting range for most models is $30<s<120\,h^{-1}$Mpc.
For the BigMD simulation the internal error estimate from the 8 octants
becomes unreliable at the largest scale so we truncate the fit at
$100\,h^{-1}$Mpc.  For the EDM and NDM models we wish to avoid fitting the
acoustic peak region entirely, since the shape of the peak is much sharper
than seen in the N-body simulations (where the peak is broadened by
non-linear evolution).  As the peak height can be lowered, in these models, by
increasing $f$ the fits become highly unstable to large $f$ if the
peak is included in the fit.  Thus we include only $s<80\,h^{-1}$Mpc for
these two models.  Note that the acoustic peak region is not very important
for this test as we are holding the cosmology and linear power spectrum fixed
and the relative error on the large scale points is much larger than those at
smaller scales.  For a more general fit information from the acoustic peak
would need to be included in some manner (e.g.~as a prior on distance).

Figures \ref{fig:bestfitPB} and \ref{fig:bestfitMD} show how well the models
fit the monopole and quadrupole moments measured from the T1 and BigMD
simulation at $z\simeq 0.55-0.56$.
(The situation is qualitatively very similar for the other mock catalogs.)
The behavior of the models near the acoustic peak has been discussed
previously.
Some of the modes which are responsible for the broadening of the acoustic
peak at $s\simeq 110\,h^{-1}$Mpc are not captured by the T1 simulations, so
the N-body peak is not as broad as it should be.  The simulation is well
converged at intermediate scales, however, and the large scales carry little
weight in the fit.
The noise from finite volume is evident in the BigMD simulation.

Note that the Eulerian models predict a higher (i.e.~more negative)
quadrupole at large scales ($40\,h^{-1}{\rm Mpc}<s<80\,h^{-1}$Mpc)
with a sharper ``break'' at small scales than the Lagrangian models,
and in particular they underpredict the N-body data on scales below
$30\,h^{-1}$Mpc.
On the tens of Mpc scales which dominate the fits the GSM and LSM agree
very well and agree quite well with the N-body data.
All of the models predict very similar behavior for the monopole over the
range $30\,h^{-1}<s<70\,h^{-1}$Mpc and in particular the N-body monopole
is well matched by all of the models in this range.

We can look for any biases or instabilities in the fit using a Monte-Carlo
Markov Chain.  We run several such chains with a parallel implementation of
the affine-invariant ensemble sampler described in \citet{GooWea10}.
To begin we allow $f$, the bias and the finger-of-god parameter to vary at
fixed linear power spectrum, with infinitely wide, flat priors\footnote{Except
for the GSM model where we imposed a prior $\sigma<10\,h^{-1}$Mpc for
numerical reasons.}.

\begin{figure}
\begin{center}
\resizebox{3.3in}{!}{\includegraphics{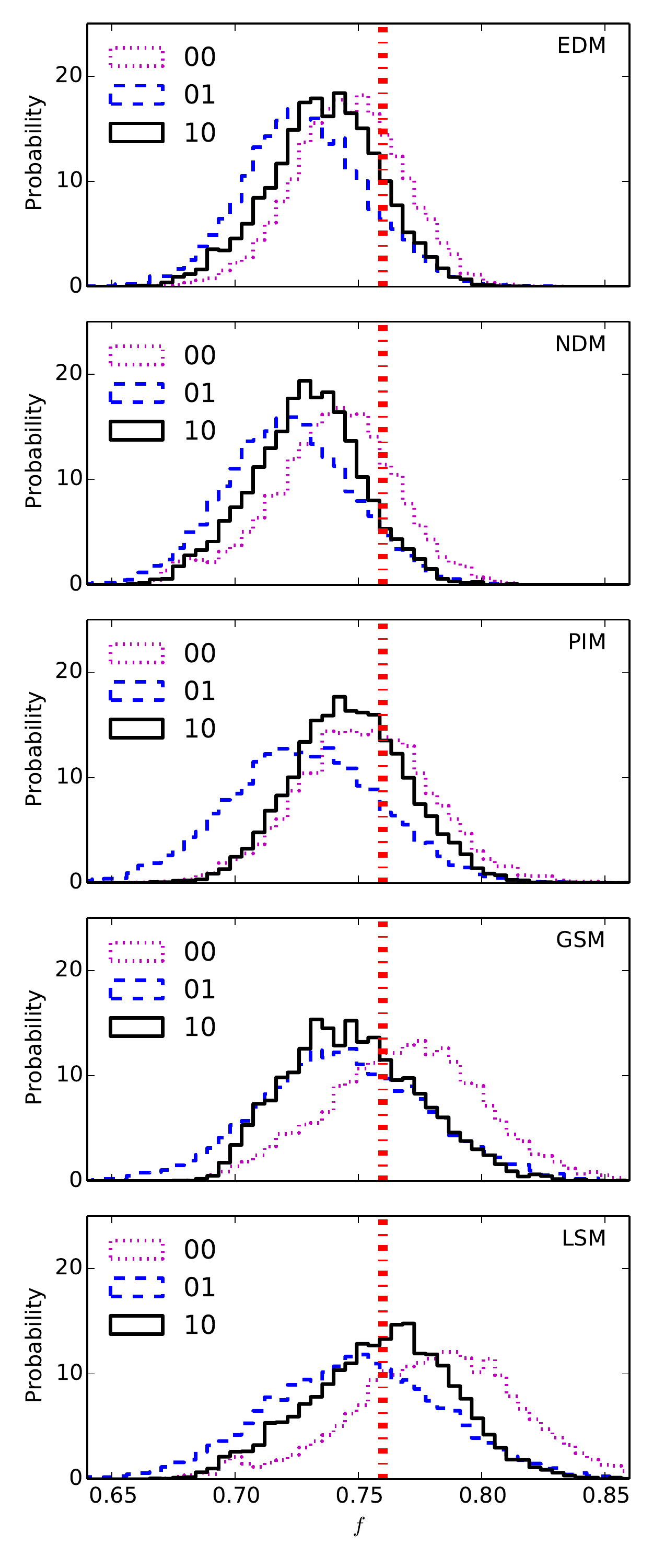}}
\caption{The marginalized posterior distribution of $f$ for models built
upon the T1 simulation, which have different prescriptions for how central
and satellite velocities are assigned and different halo finding schemes
(the numbering is described in Section \ref{sec:simulations}: 00 uses the
friends-of-friends halo center-of-mass velocity for the central galaxy, 01
enhances satellite velocities by 25 per cent and 10 is our fiducial model
with SO halos).
As in the earlier figures, the vertical line marks the value of $f$
appropriate to this cosmology at $z\simeq 0.55$ and the data were fit over
the range $30<s<120\,h^{-1}$Mpc.  As the T1 simulations cover about five
times our fiducial $5\,(h^{-1}{\rm Gpc})^3$ we expect shifts of around
$0.5\,\sigma$.}
\label{fig:fdist_tseries}
\end{center}
\end{figure}

The marginalized posterior distribution for $f$ is shown in
Fig.~\ref{fig:fdist} for the models and simulations
(Fig.~\ref{fig:fdist_tseries} highlights the T1 simulations not shown here).
The linear theory power spectrum has been held fixed and the other two
parameters, bias and finger-of-god dispersion, have been marginalized
over.  We also show the profile likelihood \citep{Wil38}, i.e.~the value of
the best fit at fixed $f$, allowing the large-scale bias and finger-of-god
parameter to vary.  This statistic provides information on the goodness of
fit which is not affected by parameter volume effects.
As we shall see, the agreement between the posterior and profile likelihood
is reasonably good in all cases indicating that our posterior likelihoods are
not unduly misleading due to volume effects, but we shall return to this
point below.

The models all fare reasonably well for all simulations, but biases are
visible at the $1\,\sigma$ level for several models.
While they fare well for the BigMD simulation, the EDM and NDM models are
biased slightly low for T0 and T1-10.
This is a consequence of the mismatch in the quadrupole (compared to the
other theories) seen in Fig.~\ref{fig:theory}.
The perturbation-inspired model performs well for all of the simulations and
the Lagrangian models do not show evidence for bias for any of these
simulations either: the distribution of $f$ values inferred from the fit
is very well centered on the correct value for most simulations.

Since we are fitting our models to correlation functions measured from finite
simulation volumes, we expect some scatter between the peak of the likelihood
and the `true' values, even in the case of a perfect model.
The size of the typical offset should roughly be
$\sqrt{V_5}\sigma_f$ where $V_5$ is the total volume of the simulations in
units of our fiducial volume, $5\,h^{-3}{\rm Gpc}^3$. 

For the T1 simulation we generated three catalogs which differ in the way
dark matter halos are identified and populated with mock galaxies
(see section \ref{sec:treepm}).
We compare the $f$ distributions in Fig.~\ref{fig:fdist_tseries}.  Again the
simpler Eulerian models tend to lie slightly low while the other models tend
to perform better.
This is quite a significant test of our simple, 1-parameter finger-of-god
prescription.  As shown in Fig.~\ref{fig:moments}, the shape of the quadrupole
at small scales is different between the catalogs and yet the value of $f$ is
well recovered in all cases.
This indicates that our simple 1-parameter finger-of-god prescription is
adequate to describe the effects of such galaxy motions on scales above
$30\,h^{-1}$Mpc.

In all of the above comparisons the posterior distribution of $f$ for the
EDM and NDM models lie low compared to the true value of $f$ in the T1
simulation.
We now examine how the fits behave as we vary the minimum scale in the fit,
$s_{\rm min}$.  Fig.~\ref{fig:fdist_rmin} shows the posterior distribution
of $f$ for the minimum $s$ of both the monopole and quadrupole varying
$s_{\rm min}$ from 20 to $40\,h^{-1}$Mpc
(the larger value is the range over which they have typically been used
\citealt{Chuang13,San13,San14}).

We see the expected behavior, an increasing bias but tighter distribution
as we lower $s_{\rm min}$.  For all models but LSM the bias at
$s_{\rm min}=20\,h^{-1}$Mpc is significant.  While LSM appears to fare well
in this test, the bias for this model can rise to $1\,\sigma$ for other
simulations in the T1 series (specifically model 01, with enhanced satellite
galaxy velocities).
Increasing $s_{\rm min}$ to $24\,h^{-1}$Mpc when fitting the LSM returns
unbiased fits for all catalogs, with the exception of T1-00.  For this
model the likelihood is broad and centered roughly $1\,\sigma$ higher than
the `true' value in the simulation.  The difference in $\chi^2$ between the
best fit value of $f$ and the `true' value is small however
($\Delta\chi^2<1$)
suggesting that the model does not rule out the correct value of $f$.
For this reason, we see no evidence that the LSM cannot be used down to
$\simeq 25\,h^{-1}$Mpc.

\begin{figure}
\begin{center}
\resizebox{3.3in}{!}{\includegraphics{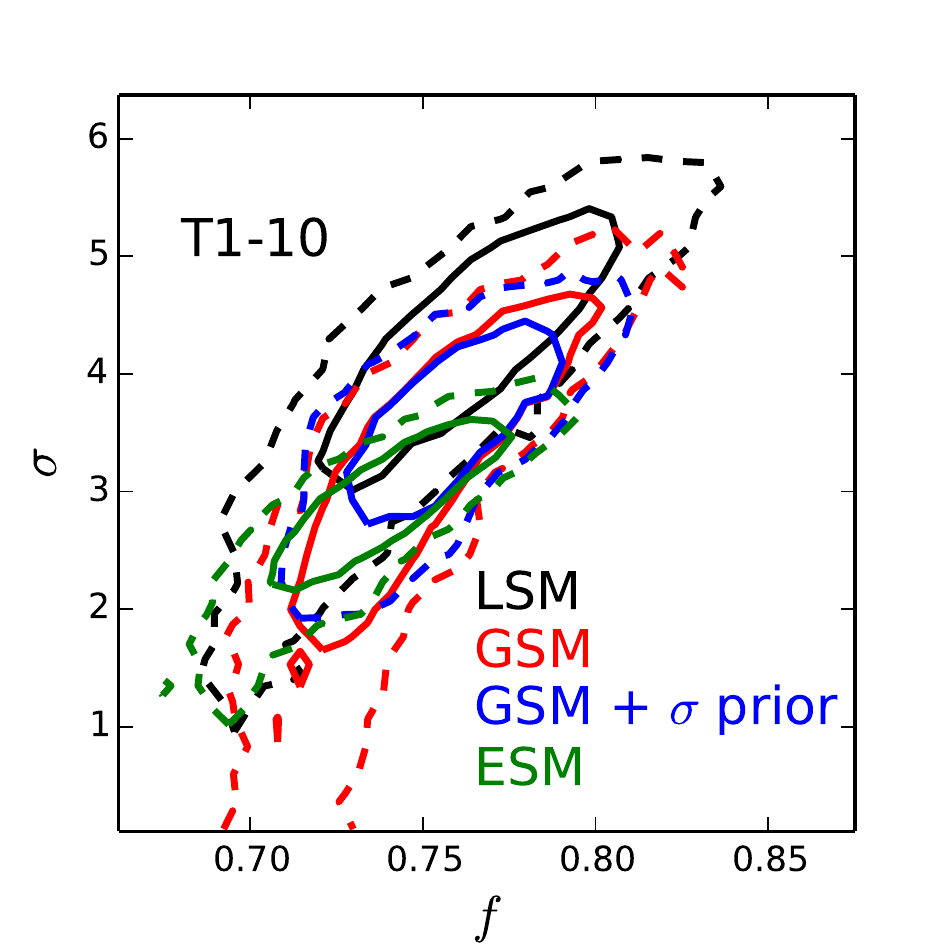}}
\caption{The marginalized joint posterior distribution of $f$ and $\sigma$ for 
the T1-10 mock for the LSM (black), GSM (red), and EDM (green) models.  In each 
case there is a positive correlation between $f$ and $\sigma$.  If we apply the 
prior on $\sigma$ advocated in \citet{Reid12} for the GSM model, low values of 
$\sigma$ are disfavored (blue).  The central value of $f$ shifts by 
$\sim 0.3\,\sigma$ and the uncertainty on $f$ is reduced by 25 per cent after
the application of the $\sigma$ prior.}
\label{fig:fsigma}
\end{center}
\end{figure}

\begin{figure}
\begin{center}
\resizebox{3.3in}{!}{\includegraphics{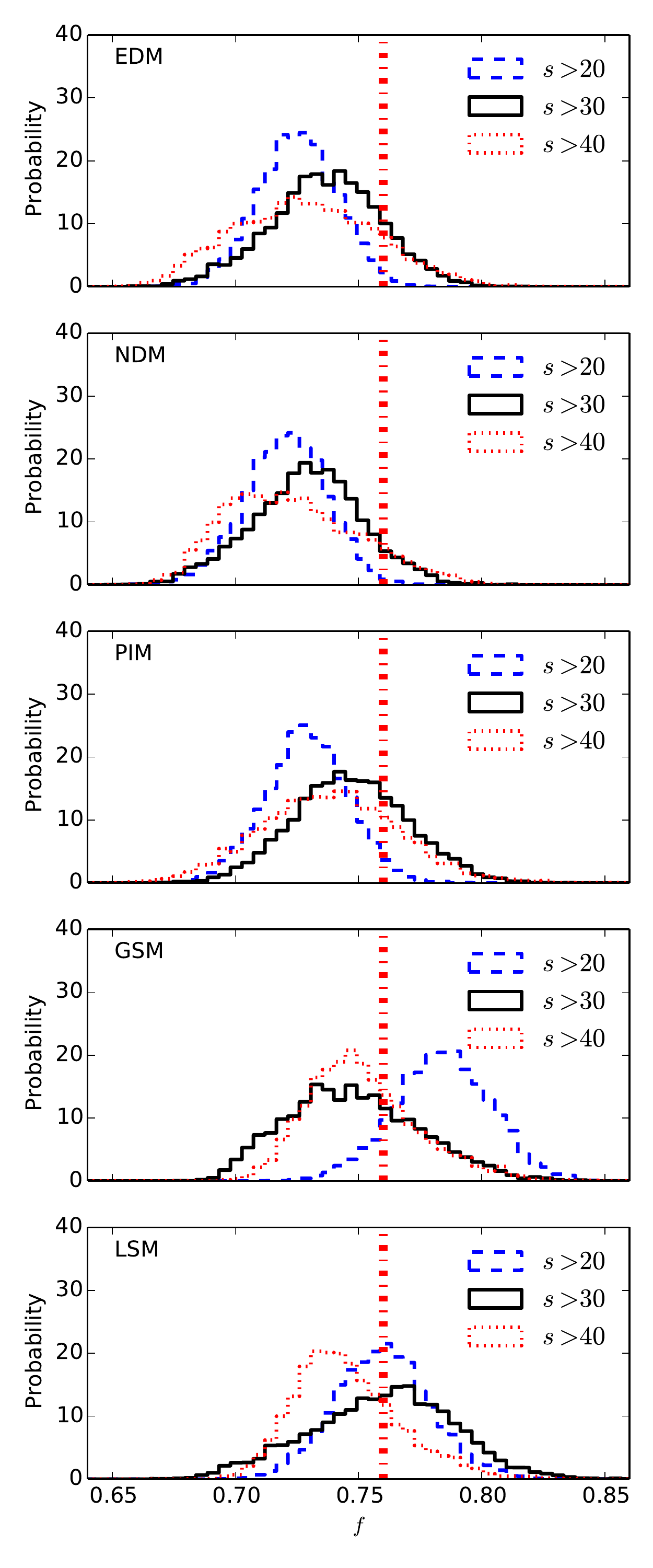}}
\caption{The marginalized posterior distribution of $f$ as a function of
the minimum scale in the fit (in $h^{-1}$Mpc) for each theoretical model.
Again the T1-10 simulation is used and the vertical line marks the value
of $f$ appropriate to this cosmology at $z\simeq 0.55$.  A similar behavior
is seen for the other simulations (see text).}
\label{fig:fdist_rmin}
\end{center}
\end{figure}

As we increase $s_{\rm min}$ we see generally less bias and a wider posterior
distribution, though the bias remains significant for EDM and NDM even at
$40\,h^{-1}$Mpc.
The behavior of the LSM is at first sight surprising, because the bias appears
to increase with increasing $s_{\rm min}$.
This can be traced to the positive correlation between $f$ and the
finger-of-god parameter, $\sigma$, shown in Fig.~\ref{fig:fsigma}
\citep[see also the discussion in][]{Reid12}.
As we increase $s_{\rm min}$ the finger-of-god parameter becomes less
constrained, and newly allowed low values of $\sigma$ correspond to lower
values of $f$.  The peak of the marginal distribution thus shifts to lower
$f$.  If we had a well motivated prior for $\sigma$ this degeneracy would
be less important.
For instance, \citet{Reid14} propose a Gaussian prior on $\sigma^2$ around
$14\pm 5$ ($h^{-1}$ Mpc)$^2$.
As shown in Fig.~\ref{fig:fsigma}, this prior disfavors low values of $\sigma$,
corresponding to models in which the galaxies have small intra-halo velocities
inconsistent with the observed anisotropic clustering on small scales.
Application of this prior shifts the central marginalized value of $f$ upwards
by $\sim 0.3\sigma$ (thus removing the small bias evident in
Figs.~\ref{fig:fdist_tseries} and \ref{fig:fdist}) and reduces its uncertainty
by 25 per cent.  Since the other mock catalogs have purposely not been tuned
to match the observed small scale anisotropic clustering, the proposed prior
is not applicable to those mock samples and we have not attempted to use it.

\begin{figure}
\begin{center}
\resizebox{3.3in}{!}{\includegraphics{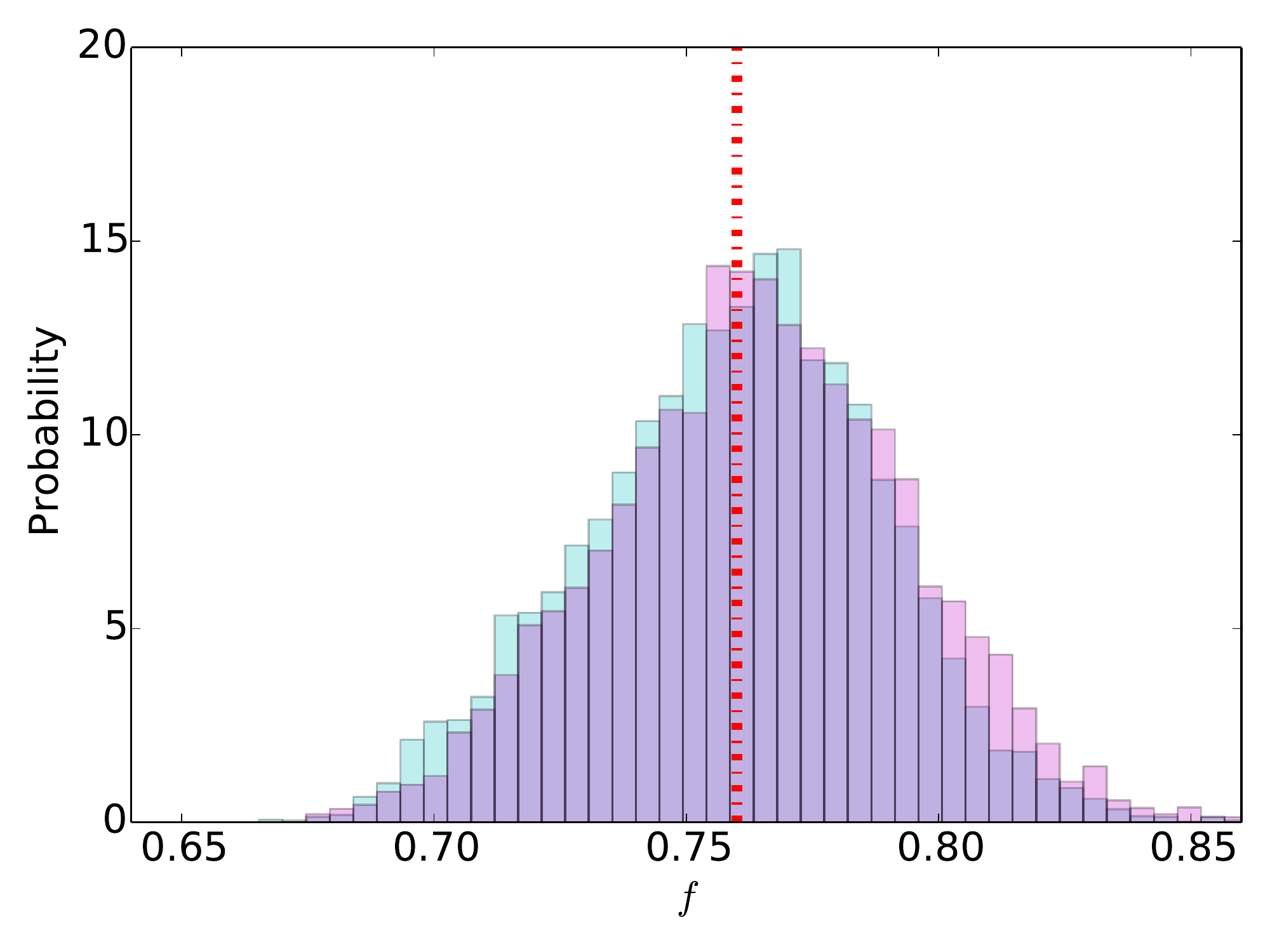}}
\caption{The marginalized posterior distribution of $f$ for the fiducial
cosmology allowing the linear theory power spectrum (and cosmological
parameters) to vary within a CMB prior (magenta histogram) compared with
the case of $P_{\rm lin}(k)$ fixed (cyan histogram).
Within the $\Lambda$CDM family current CMB data tightly constrain the
linear theory power spectrum, so that the constraints with fixed
$P_{\rm lin}(k)$ or with a CMB prior are almost identical.
Again the vertical line marks the value of $f$ appropriate to this cosmology
at $z\simeq 0.55$.}
\label{fig:fdist_cmb}
\end{center}
\end{figure}

The tests presented above are quite stringent, as they allow only a
minimum number of parameters to vary.  It is also of interest to ask how
well the models perform when we allow the linear theory power spectrum
and cosmological parameters to vary.
In Fig.~\ref{fig:fdist_cmb} we show the marginalized posterior for $f$ from
the LSM model, allowing the linear theory power spectrum to vary within a CMB
prior as in \citet{Samushia14}.  Specifically we allow
$\omega_c\equiv\Omega_c h^2$, $\omega_b\equiv\Omega_b h^2$, $n_s$, $f$
$\sigma$ and the large-scale bias to vary with a prior on $\omega_c$,
$\omega_b$ and $n_s$.  We hold $h$ and $\sigma_8$ fixed at their fiducial
values.  As discussed extensively in \citet{Reid12,Samushia13,Samushia14}
within the $\Lambda$CDM family current CMB data tightly constrain the
linear theory power spectrum so that allowing it to float or holding it fixed
return almost the same marginalized constraints on $f$.
This is clearly illustrated by Fig.~\ref{fig:fdist_cmb}.

\subsection{Observational effects}

The above tests have been performed on ``ideal'' measurements, involving
constant time outputs, with constant number density done in periodic boxes.
We now turn to a discussion of how departures from this ideal situation
affect measurements of the growth rate.

In this section we make use of the QPM mock catalogs, in a differential
fashion.  We compare the $\xi_\ell$ measured on the constant time, uniform
density, periodic QPM boxes to that computed on catalogs which have been
projected onto the sky and observed under various constraints.
We shall call these latter catalogs sky mocks.
Throughout we shall work with the average of 100 sky mocks, sufficient to
look for effects at much less than the expected statistical error of BOSS.
We now show that the effects on $\xi_\ell$ (and the covariance matrix) at
the relevant scales are unimportant.

First we compare the average correlation function of the sky mocks to that
of the periodic mocks, assuming no fiber collisions, redshift failures or
systematics.  The major differences are from the shape of $\bar{n}(z)$
and (to a lesser extent) wide-angle redshift-space distortions.
We convert angles and redshifts to distances using the proper cosmology for
the simulations, use the correlation function estimator of \citet{LanSza93}
and assign redshifts to random points by `shuffling' the data redshifts,
which is the technique used in most of the BOSS papers to date.
We use twenty times as many randoms as data points.

The average $\xi_\ell$ from the sky and periodic mocks differ by a small
but measurable amount, even on large scales.  This major driver of this
difference is the redshift distribution chosen for the random points.
This has been studied before, most recently by \citet{SamPerRac12}.  Those
authors showed that our approach gives the smallest error for a wide-angle
survey like BOSS, but this error is not negligible.
Luckily the error is very small (less than 1 per cent) on the smallest scales
and grows more slowly than the statistical error as we move to larger scales.
Since our measurements are heavily weighted to small scales this effect is
actually less important for redshift-space distortions than for
e.g.~anisotropic fitting of the baryon acoustic oscillation peak.
Over the range $20-60\,h^{-1}$Mpc the effect is always less than $2.5$ per cent,
in both the monopole and quadrupole moments of the redshift-space correlation
function.
This is much less that our assumed statistical errors and comparable to the
systematics from the modeling.

Next we introduce `fiber collisions'.  To mimic this observational effect
we divide the mock galaxies into those which have a neighbour within
$62^{\prime\prime}$ and those which do not.
We measure from the BOSS sample, as a function of position on the sky,
the fraction of close pairs where one galaxy precluded the other getting a
redshift.  For each mock pair at the same sky position we randomly remove
one galaxy from the pair, increasing the weight of its neighbor by one.
If a galaxy has more than one near neigbhor we choose one at random.
This is only an approximation to the fiber assignment scheme employed by
BOSS on the real data \citep{Tiling}, but it captures the main features of
the effect.

In common with earlier work \citep[see e.g.][for a recent study]{Reid14}
we find that fiber collision correction using nearest neighbors has little
impact on the correlation function on large scales.
The monopole of the redshift-space correlation function is affected by
less than one per cent on scales above $10\,h^{-1}$Mpc while the quadrupole
is affected by less than one per cent above $30\,h^{-1}$Mpc and only by two
per cent at $20\,h^{-1}$Mpc.  Both of these errors are comfortably below the
statistical and modeling error and can be ignored.

Next we discuss the impact of evolution across the sample, and the
approximation inherent in interpreting our measurement as $\xi_\ell$
`at' a given redshift.

\subsection{Lightcone evolution}

If the number density is slowly varying with redshift over the scales which
dominate our fits\footnote{We have explicitly checked that this is true for
the BOSS DR12 data set by counting galaxy pairs and comparing to
Eq.~(\ref{eqn:zevoln}).}, the correlation function we measure is
\citep{Mat97,WhiMarCoh08}
\begin{equation}
  \xi_{\ell,{\rm obs}}(s) \simeq
  \frac{\int dz\ (dN/dz)^2(H/\chi^2)\xi_\ell(s,z)}{\int dz\ (dN/dz)^2(H/\chi^2)}
\label{eqn:zevoln}
\end{equation}
where $dN/dz$ is the redshift distribution of the sample, $H$ is the Hubble
parameter at redshift $z$ and $\chi$ is the comoving angular diameter distance
to redshift $z$.

\begin{figure}
\begin{center}
\resizebox{3.3in}{!}{\includegraphics{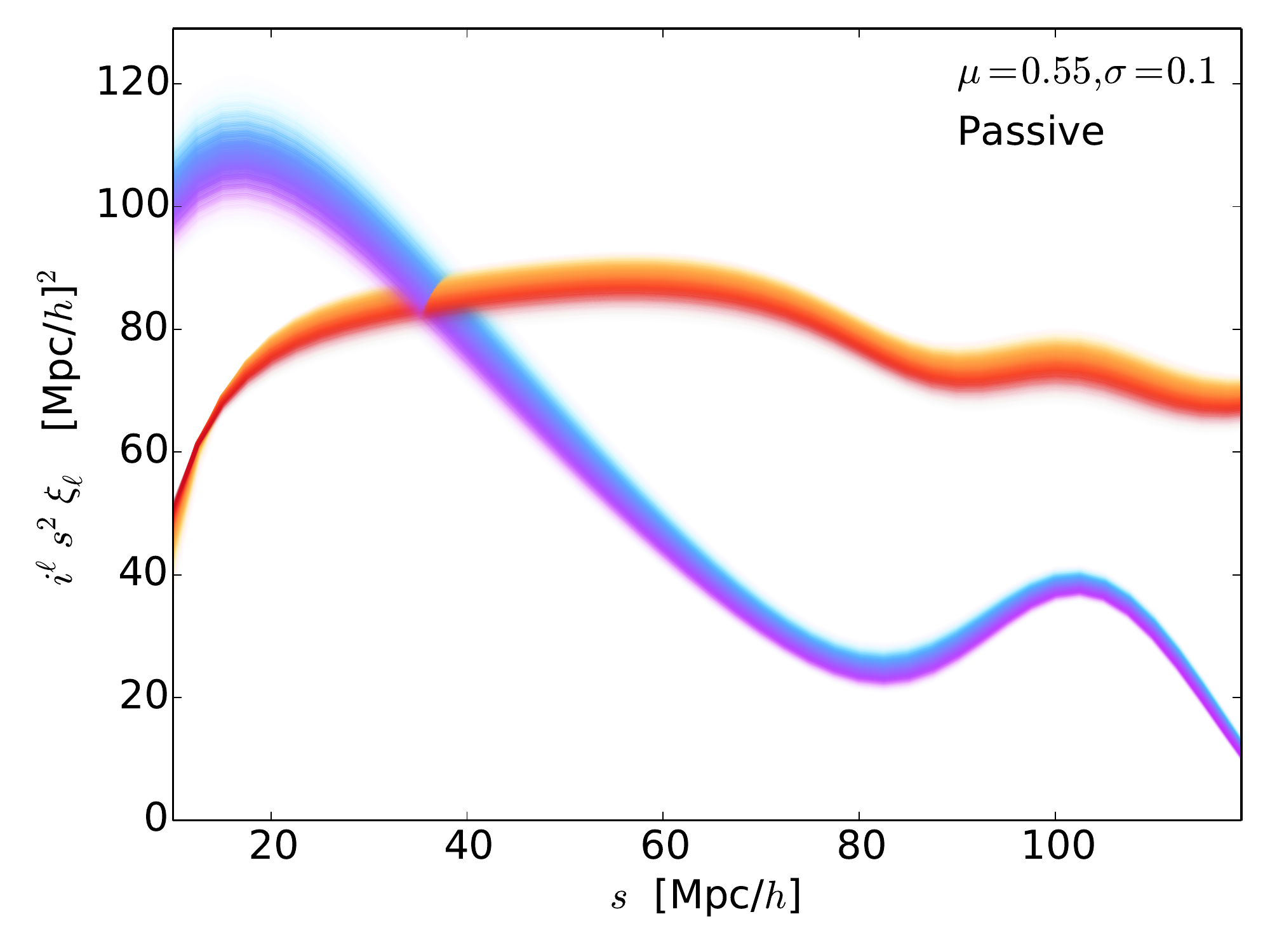}}
\resizebox{3.3in}{!}{\includegraphics{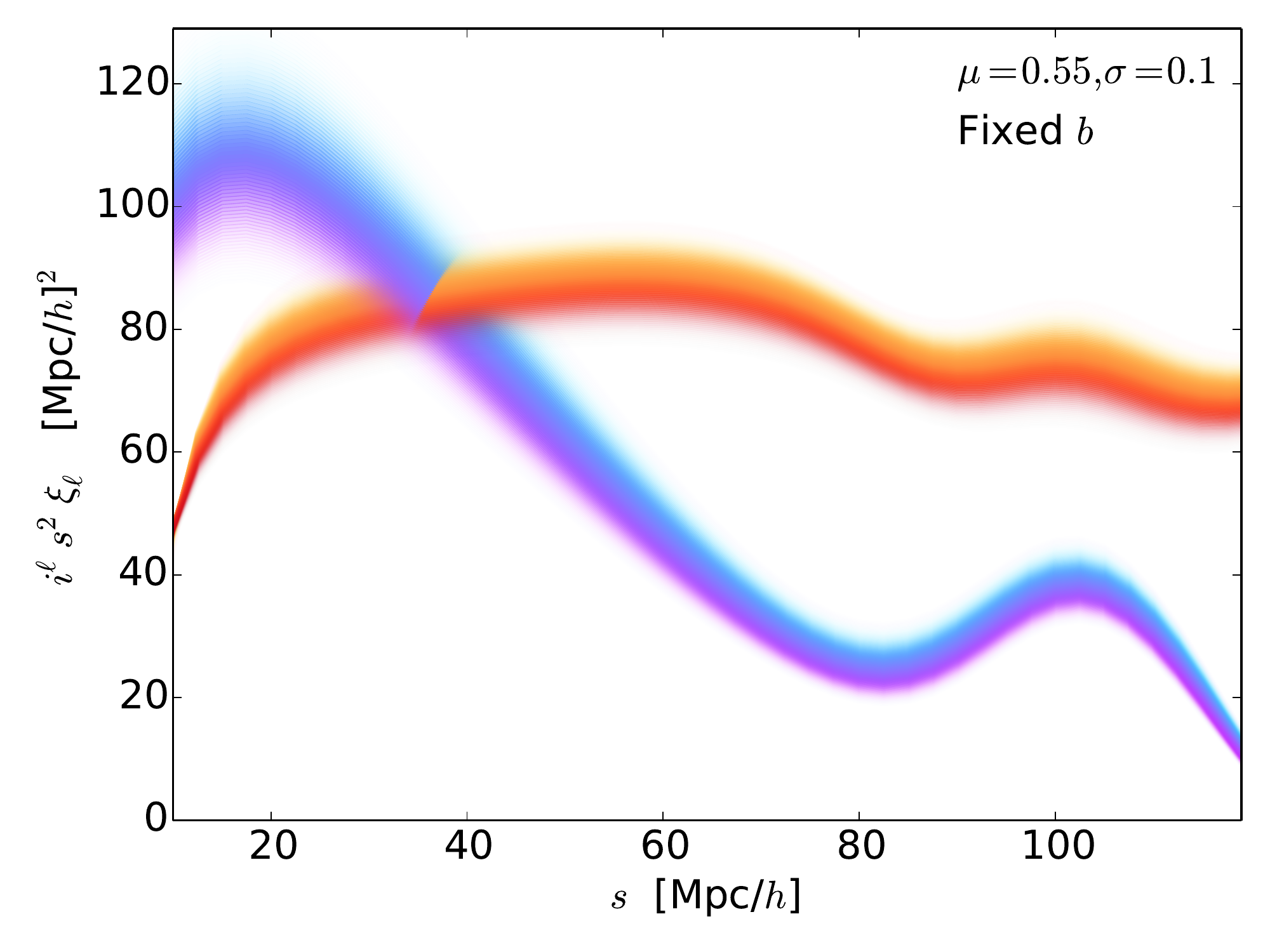}}
\caption{The evolution of the correlation function multipoles with redshift,
as determined by the LSM for the cosmology of the BigMD simulation.
In each panel the color scale indicates the redshift while the transparency
scales with the weight in Eq.~(\ref{eqn:zevoln}).
The upper panel shows the result for a Gaussian $dN/dz$ with the indicated
mean and dispersion assuming passive evolution
($b-1\propto D^{-1}$; \citealt{Fry96})
while the lower panel assumes fixed $b$.  The former is a better approximation
to the evolution of the clustering of BOSS galaxies \citep{Whi11}.}
\label{fig:zevoln}
\end{center}
\end{figure}

\begin{figure}
\begin{center}
\resizebox{3.3in}{!}{\includegraphics{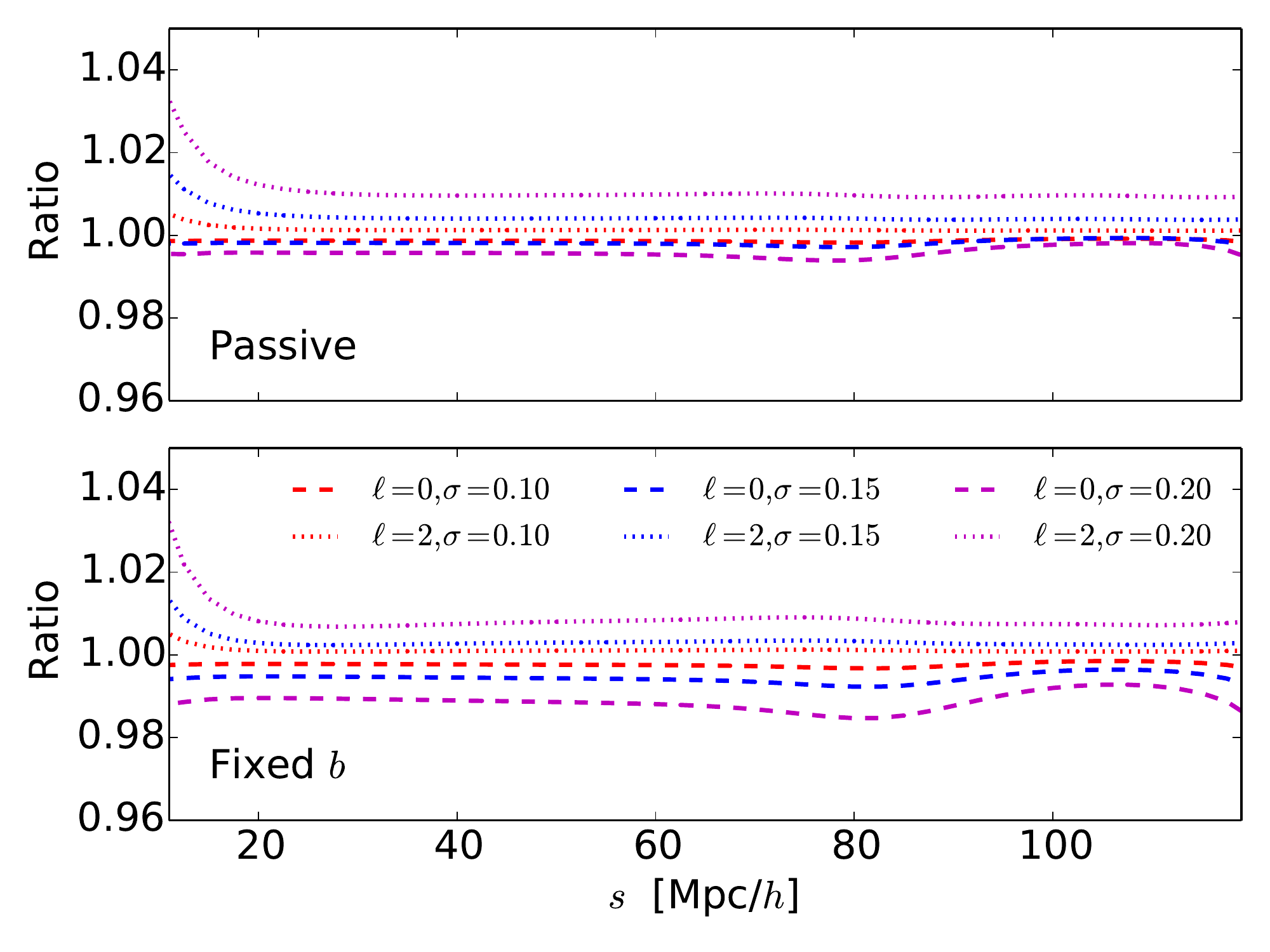}}
\caption{The ratio of the redshift-averaged correlation function multipoles
to the model evaluated at $z_{\rm eff}$, i.e.~using a constant-time
approximation, for Gaussian $dN/dz$ with the listed mean and standard
deviation (for the $\sigma=0.2$ case we have set $dN/dz=0$ for $z<0.05$).
In all cases the ratio is nearly scale independent and close to unity.}
\label{fig:zevoln_ratio}
\end{center}
\end{figure}

To get a sense for the amount by which evolution could change the shape of
the correlation functions we assumed a Gaussian $dN/dz$, centered at
$z=0.55$ and with varying widths.  Using the LSM as ``truth'', and the
cosmology of the BigMD simulation, we computed the integral and compared
it to $\xi(r,z_{\rm eff})$ where $z_{\rm eff}$ is evaluated by an integral
similar to Eq.~(\ref{eqn:zevoln}) but with $z$ in the numerator rather than
$\xi(s,z)$.  Figure \ref{fig:zevoln} shows the range of values of the moments
of the correlation function take on across the redshift range for two
different assumptions about the bias evolution: passive evolution \citep{Fry96}
and constant bias.  We expect the former to be a better approximation for
BOSS galaxies, and we see predicts more constant clustering with redshift.
Figure \ref{fig:zevoln_ratio} compares the average correlation function
multipoles to those evaluated at $z_{\rm eff}$.
The ratio is very close to scale-invariant, and within a few percent of
unity even for quite broad redshift windows (for comparison the $dN/dz$
for the BOSS CMASS sample has approximately ten percent width).
This suggests that we can safely interpret the measurements of the correlation
function as at an `effective' redshift with little impact on the cosmology.

\section{Discussion}
\label{sec:discussion}

A key probe of cosmological expansion and our theory of gravity is the
growth of large-scale structure, as revealed in the clustering
of galaxies observed in large redshift surveys.  A growing understanding
of the process of structure formation allows us to compare well-motivated
models to data from ever larger surveys to provide tight constraints on
deviations from the standard model of cosmology.  In this paper we have
compared the predictions from a number of phenomenological models for the
anisotropic clustering of galaxies measured in configuration-space against
a series of mock catalogs designed to mimic the high redshift sample from
BOSS.

Even dealing with `ideal' data, we have found that none of the proposed
models fit the mock data on all scales, but they all do a good job of
fitting the monopole and quadrupole moments of $\xi$ in the range
$40<s<80\,h^{-1}$Mpc.  Dispersion models based on the simplest versions
of Eulerian perturbation theory do not provide a good fit to the baryon
acoustic peak feature of BOSS-like galaxies at $z\simeq 0.5$, but fare
well on the shape of the monopole and quadrupole on intermediate scales.
They tend to slightly underpredict the growth rate of structure when fit
down to $20-30\,h^{-1}$Mpc scales.

More sophisticated models, and models based on Lagrangian perturbation theory,
such as the GSM and LSM (Section \ref{sec:models}), provide unbiased fits to
the data on scales above $25-30\,h^{-1}$Mpc.
In particular the LSM is straightforward and fast to compute and provides
a reliable model for fitting BOSS-like galaxies at $z\simeq 0.5$.

Including non-idealities such as fiber collisions or lightcone evolution
does not appreciably modify these conclusions.  Using standard corrections
for fiber collisions and interpreting the light-cone data as a measurement
at an effective redshift introduces errors which are small compared to other
sources of uncertainty.

\vspace{0.2in}
M.W. would like to thank Chris Blake for helpful comments on an early
draft of the manuscript.
M.W. is supported by NASA.
This work made extensive use of the NASA Astrophysics Data System and
of the {\tt astro-ph} preprint archive at {\tt arXiv.org}.
The analysis made use of the computing resources of the National Energy
Research Scientific Computing Center.
The MultiDark Database used in this paper and the web application providing
online access to it were constructed as part of the activities of the
German Astrophysical Virtual Observatory as result of a collaboration between
the Leibniz-Institute for Astrophysics Potsdam (AIP) and the Spanish MultiDark
Consolider Project CSD2009-00064.
The BigMD simulation suite was performed in the supercomputer at LRZ.


\newcommand{\aj}{AJ}
\newcommand{\apj}{ApJ}
\newcommand{\apjs}{ApJ Suppl.}
\newcommand{\araa}{ARA{\&}A}
\newcommand{\aap}{A{\&}A}
\newcommand{\pre}{PRE}
\newcommand{\prd}{Phys. Rev. D}
\newcommand{\apjl}{ApJL}
\newcommand{\physrep}{Physics Reports}
\newcommand{\nat}{Nature}

\setlength{\bibhang}{2.0em}
\setlength\labelwidth{0.0em}
\bibliography{biblio}

\label{lastpage}
\end{document}